\documentclass[11pt]{article}

\usepackage{booktabs}
\usepackage[english]{babel}
\usepackage{amsmath,amssymb,amsbsy,amstext, amsthm, simplewick}
\usepackage{hyperref}
\usepackage{graphicx}
\usepackage{amsfonts}
\usepackage{amssymb}
\usepackage[small]{caption}
\usepackage{upgreek}
\usepackage{cancel}
\usepackage{color}
\usepackage[dvipsnames]{xcolor}
\usepackage{subcaption}

\usepackage{colortbl}
\definecolor{lightgreen}{cmyk}{0.2, 0, 0.2, 0.2}
\definecolor{lightgray}{cmyk}{0.1,0.2,0,0.1}
\definecolor{lightgray2}{cmyk}{0.1,0.1,0,0.1}

\makeatletter
\newlength{\apb@width}
\newcommand{\autoparbox}[2][c]{\settowidth{\apb@width}{#2}\parbox[#1]{\apb@width}{#2}}

\makeatother

\numberwithin{equation}{section}

\def\beq{\begin{equation}}
\def\eeq{\end{equation}}

\def\bea{\begin{eqnarray}}
\def\eea{\end{eqnarray}}

\def\beq{\begin{equation}}
\def\eeq{\end{equation}}
\def\bea{\begin{eqnarray}}
\def\eea{\end{eqnarray}}

\DeclareRobustCommand{\SkipTocEntry}[4]{}

\newcommand{\edit}[2]{\ignorespaces}

\newcommand{\appropto}{\mathrel{\vcenter{
  \offinterlineskip\halign{\hfil$##$\cr
    \propto\cr\noalign{\kern2pt}\sim\cr\noalign{\kern-2pt}}}}}

\begin{document}

\begin{titlepage}

\setcounter{page}{1} \baselineskip=15.5pt \thispagestyle{empty}
\begin{flushright}
\end{flushright}

\bigskip
\vspace{1.2cm}
\begin{center}
\vspace{0.1cm} {\fontsize{20}{30}\selectfont  \bf A Case for Future Lepton Colliders}
\end{center}

\vspace{0.5cm}
\begin{center}
{\fontsize{14}{30}\selectfont   Nathaniel Craig}
\end{center}

\vspace{0.2cm}

\begin{center}
\vskip 8pt
\textsl{Department of Physics, University of California, Santa Barbara, CA 93106, USA}
\end{center}

\vspace{1.2cm}
\hrule \vspace{0.3cm}
{ \noindent {\sffamily \bfseries Abstract} \\[0.1cm]
I present a case for proposed future linear and circular lepton colliders as ideal machines for understanding and exploration. As machines for understanding, they provide unprecedented tools for studying the Higgs and observing phenomena never before seen in nature. As machines for exploration, they are ideally suited to discovering new particles interacting too weakly to be seen otherwise. These arguments for future lepton colliders are largely independent of the as-yet-unknown outcome of the LHC physics program. \\

\noindent Based on a public plenary and an invited talk presented at the International Workshop on Future Linear Colliders (LCWS2016), Morioka, Japan, 5-9 December 2016. C16-12-05.4.
\noindent}
\vspace{0.3cm}
\hrule

\vspace{0.6cm}

\end{titlepage}

\newpage 

\section{Introduction}

\begin{figure}[t]    \centering
   \includegraphics[width=5in]{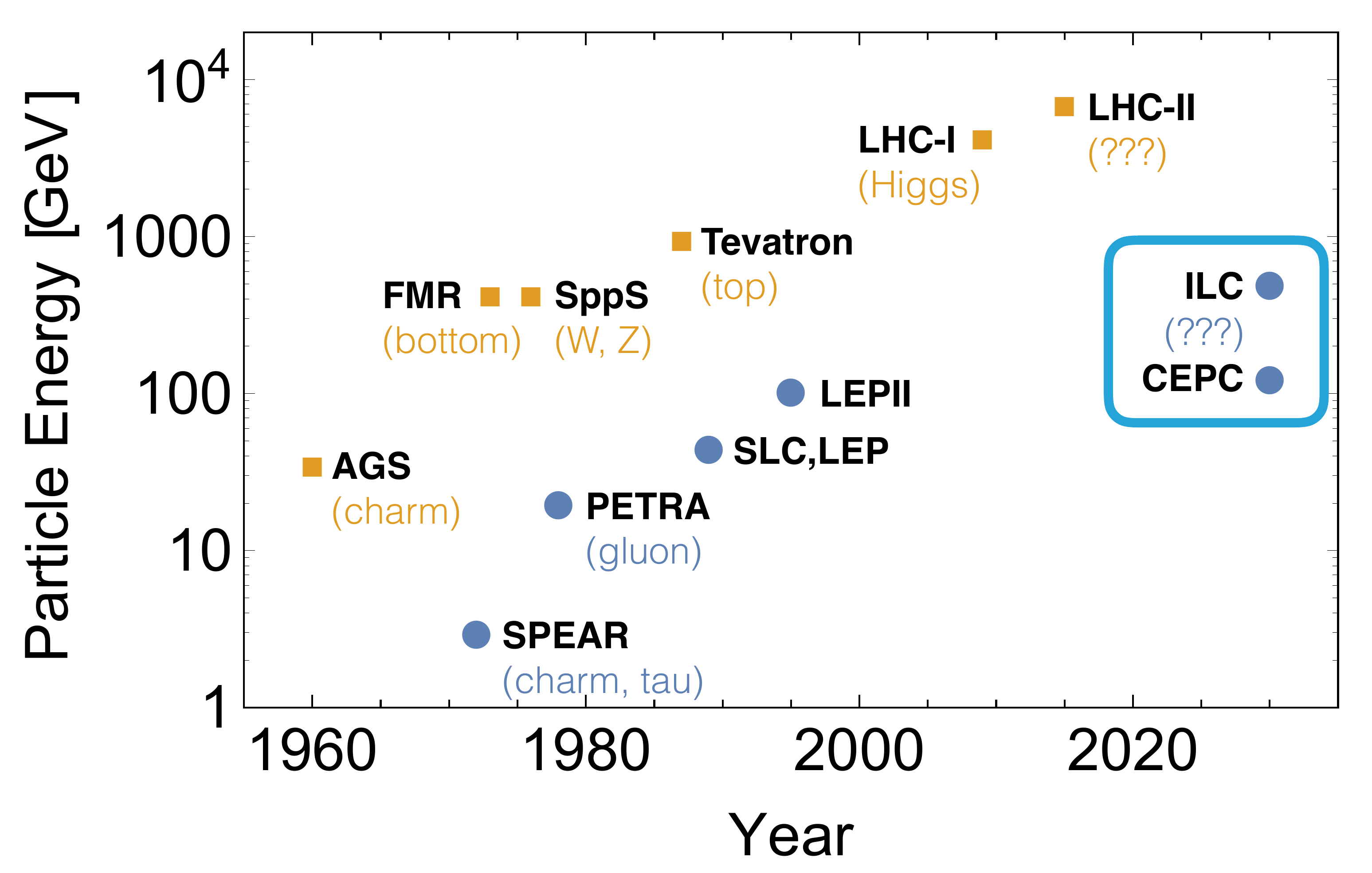} 
   \caption{A selective Livingston plot illustrating various accelerator complexes (hadrons in orange, leptons in blue) and the new fundamental particles discovered therein. While SLC and LEP did not directly discover new particles, their precision measurements telegraphed the mass scale of the Higgs, subsequently discovered at the LHC.}
   \label{fig:livingston}
\end{figure}

Particle physics is about addressing fundamental questions -- Where did we come from? Where are we going? What are we made of? How does it all work? We typically approach these questions by seeking to better understand the things we see around us, and by exploring the unknown to reveal things we have not yet seen. While we have a variety of tools at hand to pursue these goals, particle accelerators are particularly powerful means of furthering understanding and exploration. Among other things, the discovery and exploration of the Standard Model has proceeded over the course of the last half-century thanks to a series of increasingly powerful lepton and hadron colliders. As illustrated in Figure \ref{fig:livingston}, more or less every advance in energy or precision leading to the discovery of a new fundamental particle. Equally important are the precision measurements made at accelerators such as SLC and LEP, which gave strong indication of the mass range in which the Higgs was eventually discovered and outlined precise targets for future machines of discovery. 

In the present era we are fortunate to have an extraordinary tool in the form of the LHC, which has already discovered the Higgs boson and will powerfully probe unknown territory throughout the remainder of its lifetime. That said, given the timescales involved in the design and construction of colliders, it's useful to think carefully about what machines might come next, given what we know thus far. Of course, this exercise is complicated by the fact that we don't know what lies in store at the LHC -- we may discover entirely unforeseen new particles, or we may provide beautiful validation of the Standard Model to the TeV scale, or we may end up somewhere in between. It is possible, and reasonable, to frame arguments for future lepton and hadron colliders in terms of the potential outcomes of the LHC. However, these arguments are necessarily contingent, and contingent arguments often lack the sort of momentum that it takes to carry the day. 

Rather than worrying about contingencies, however, a strong argument can be formulated without reference to the possible outcomes of the LHC. There are profound questions at hand that the LHC is not ideally suited to answer, and endeavoring to answer these questions provides sharp motivation for various future accelerators. My goal here is to articulate some of these questions, and provide some qualitative and quantitative arguments for the tremendous potential of future lepton colliders to furnish the answers. I should note that the arguments presented here are not particularly original, and aspects of them may be found in the excellent design reports for various proposed lepton colliders (e.g., \cite{Linssen:2012hp, Baer:2013cma, Gomez-Ceballos:2013zzn, CEPCPreCDRvolumn2}).
Nonetheless, I hope to provide a bit of fresh perspective, as well as a few new quantitative results.

Before we begin in earnest, some caveats: for the most part, when it comes to details I will focus on the parameters of the proposed Circular Electron-Positron Collider (CEPC) and the International Linear Collider (ILC). That is not to discount the comparable (and often greater) promise of proposed lepton colliders such as FCC-ee and CLiC, but because the timescales for the CEPC and ILC are potentially somewhat shorter, and the planned reach somewhat more conservative, they provide a useful benchmark for framing quantitative arguments. Likewise, there are sharp arguments for the qualitatively 
new questions answerable by future hadron colliders operating at 50-100 TeV (beyond the expected incremental improvements that come from increased energy), and my focus on lepton colliders is not intended to diminish their appeal.

\section{Lepton Colliders as Machines for Understanding}

Let's begin with some arguments for future lepton colliders as machines of understanding. This is, of course, the more familiar context for motivating lepton colliders; the extraordinary precision of past lepton colliders has been instrumental in testing and understanding the framework of the Standard Model, and it is not unreasonable to expect that lepton colliders may play a comparable role in the future. As we look to the future, there are particularly sharp places where the precision of lepton colliders can provide qualitatively new insight into phenomena we have already observed.

The Higgs discovery at the LHC was one of the great triumphs of 21st-century physics, and marked the start of a comprehensive program for studying Higgs properties. But for all the excitement of discovery, we still know very little about the Higgs.  There are fundamental questions about the Higgs boson that are essentially un-answerable at the LHC, and for which lepton colliders are ideally suited. Here are two that are particularly compelling:

\begin{itemize}
\item The Higgs appears to be a scalar, a particle without intrinsic spin. We've seen examples of spinless composite scalars and pseudoscalars before, as a result of confinement in QCD, but we've never seen an elementary scalar. Is the Higgs something familiar, or entirely new? The sharp question is, {\it is the Higgs elementary or composite?} Another way to put this would be to ask if the Higgs is exactly point-like, as expected of an elementary scalar, or if it possesses some sort of size, as expected of a composite scalar. Ultimately the best we can do is bound the effective scale of compositeness, so a good way to frame the question is in terms of the ``size'' of the Higgs relative to its (reduced) Compton wavelength, $\sim 1/m_h$ in natural units. Of the composite scalars and pseudoscalars observed so far in nature, the most pointlike are those such as the $\pi^0$ and $\pi^\pm$ whose size is significantly smaller than their Compton wavelength -- by about a factor of six if we take the size of the pions to be set by the $\rho$ meson.

\item The Standard Model predicts that the Higgs interacts with itself (without changing any quantum numbers), unlike any other particle we've yet seen in nature. Here the sharp question is, {\it does the Higgs interact with itself?} That is to say, does the Higgs have a trilinear coupling, and does its magnitude agree with Standard Model predictions? 
\end{itemize}

While both of these questions can be asked at the LHC (and will continue to be asked throughout the entirety of its existence), the precision of available answers is fundamentally limited. In contrast, lepton colliders provide the precision necessary to answer these questions on a qualitatively new level.

\subsection{Measuring the size of the Higgs}

We can make the first point a bit more quantitative. One sharp way to parameterize the size of the Higgs, given that a purely elementary Higgs is point-like, is through the size of higher-dimensional operators characterizing the departure from purely Standard Model expectations. As far as the size of the Higgs is concerned, the key operator is 
\begin{equation}
\mathcal{O}_H = \frac{1}{2} \left(\partial_\mu |H|^2 \right)^2 
\end{equation}
An irrelevant operator, it comes with a coefficient $c_H$ and a scale $\Lambda$, extending the Standard Model lagrangian as $\delta \mathcal{L} = \frac{c_H}{\Lambda^2} \mathcal{O}_H$. There are various senses in which this can be thought of as providing the leading parameterization of the compositeness of the Higgs. Among other things, this operator is generated purely by the strongly-interacting sector in the case of a strongly-interacting light higgs (SILH) \cite{Giudice:2007fh}, and parametrically represents the leading interaction indicative of composite origin of the Higgs.

The leading effect of this operator, after electroweak symmetry breaking, is to alter the normalization of the kinetic term of the physical Higgs,
\begin{equation}
\frac{c_H}{\Lambda^2} \cdot \frac{1}{2}  \left(\partial_\mu |H|^2 \right)^2  \rightarrow \left( \frac{2 c_H v^2}{\Lambda^2} \right) \cdot \frac{1}{2} \left(\partial_\mu h \right)^2
\end{equation}
Canonically normalizing the Higgs by the field redefinition $h \rightarrow (1 - c_H v^2 / \Lambda^2) h$ then shifts all Higgs couplings uniformly. This operator is difficult to measure at the LHC since it shifts all couplings by the same amount, and therefore drops out of branching ratios; it can only be measured directly in the total Higgs width (which is difficult to measure at the LHC) or the production rates (whose measurements have relatively large systematic errors, which cannot be reduced in this case by measuring ratios of production rates).  The achievable precision at the LHC should eventually be on the order of $\Lambda \gtrsim 650$ GeV, corresponding to a bound on the ``Higgs radius'' that is only about a factor of five below the Compton wavelength of the Higgs. This amounts to a test of Higgs compositeness on par with the pion. 

\begin{figure}[t]    \centering
   \includegraphics[width=5in]{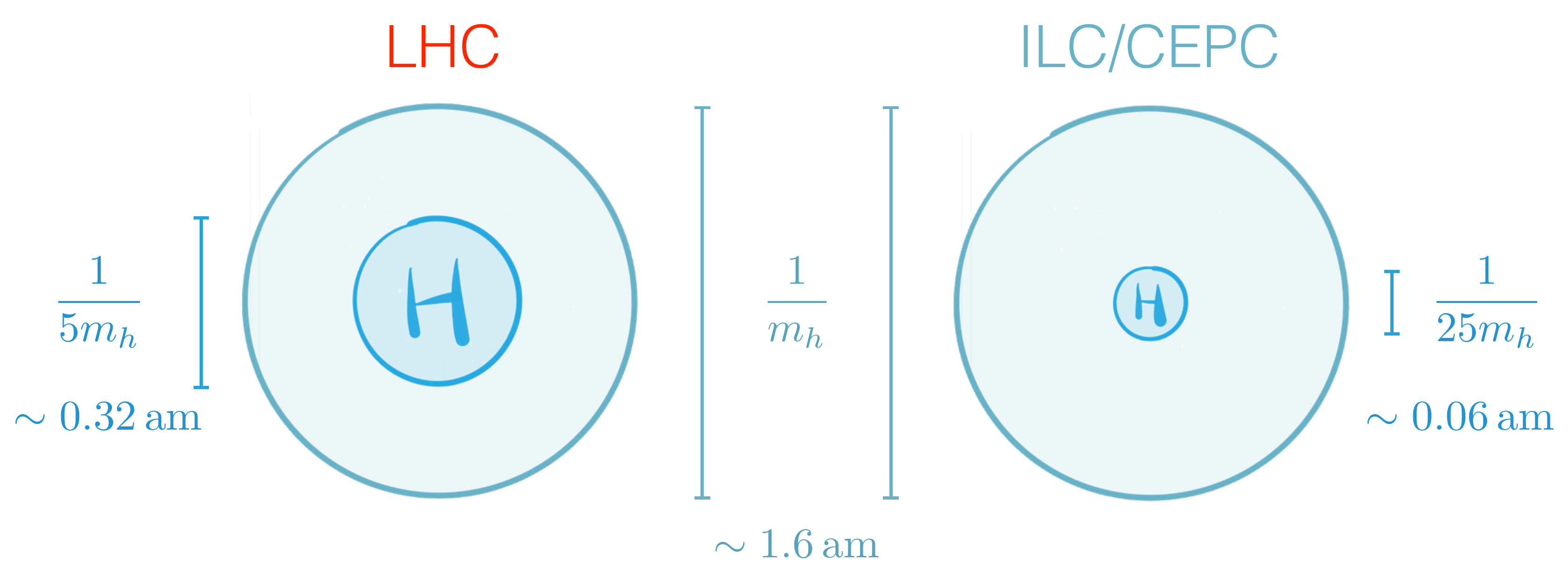} 
   \caption{A cartoon of the achievable precision in measuring the ``size'' of the Higgs at the LHC and future lepton colliders relative to the Compton wavelength of the Higgs. Not particularly to scale.}
   \label{fig:radius}
\end{figure}

In contrast, lepton colliders provide an ideal setting for constraining $c_H$, insofar as they allow a direct and model-independent measurement of the $Zh$ production cross section via $Z$ recoils. The considerable precision achieved in this channel is the cornerstone for Higgs coupling measurements at lepton colliders. Both linear and circular lepton colliders provide exceptional reach. Direct measurement of $Zh$ with a precision $\delta \sigma_{Zh} / \sigma_{Zh} = 0.88\%$ \cite{Yan:2016xyx} at the ILC leads to an anticipated $1\sigma$ bound of 
\begin{equation}
c_H \frac{v^2}{\Lambda^2} < 0.044
\end{equation}
which for $c_H = 1$ corresponds to a scale $\Lambda > 2.6$ TeV. Phrased in terms of a bound on the ``Higgs radius'', this corresponds to $r_H < 0.076$ am, or a factor of twenty below the Compton wavelength of the Higgs. At the CEPC, a measurement of $\delta \sigma_{Zh} / \sigma_{Zh} = 0.51\%$ \cite{CEPCPreCDRvolumn2} would lead to a bound $\Lambda > 3.5$ TeV and a corresponding Higgs radius $r_H < 0.056$ am -- nearly a factor of thirty below the Compton wavelength of the Higgs. The considerable improvement achievable by lepton colliders is sketched in cartoon form in Figure \ref{fig:radius}

In this respect, the appeal of lepton colliders is clear. The LHC can probe the compositeness of the Higgs to about the same fractional level of compositeness as the charged and neutral pions of QCD. Probing the elementary or composite nature of the Higgs in a quantitatively new way requires the precision of lepton colliders.

\subsection{Measuring the self-interaction of the Higgs}

We can likewise be a bit more quantitative about the probe of Higgs self-interactions. The Standard Model Higgs trilinear coupling is fixed in terms of the Higgs mass and vacuum expectation in the Standard Model, so a useful way to parameterize deviations (ranging from a vanishing self-coupling to modest departures from Standard Model expectations) is in terms of the irrelevant operator
\begin{equation}
\mathcal{O}_6 = |H|^6
\end{equation}
and with a dimensionless coefficient $c_6$ and scale $\Lambda$, it extends the Standard Model lagrangian via $\delta \mathcal{L} = \frac{c_6}{\Lambda^2} \mathcal{O}_6$. Upon electroweak symmetry breaking, this leads to deviations in the Higgs trilinear coupling relative to Standard Model expectations. Taking care to define the Standard Model prediction in terms of measured input quantities $m_H$ and $v$, the deviation due to $\mathcal{O}_6$ takes the form
\begin{equation}
\frac{m_H^2}{2 \sqrt{2} v} h^3 \rightarrow \frac{m_H^2}{2 \sqrt{2} v} \left( 1 + 8 \frac{v^2}{m_H^2} \frac{v^2}{\Lambda^2} c_6 \right) h^3
\end{equation}
 and a measurement of the Higgs self-coupling can be interpreted as a bound on $c_6$. Measurement of the Higgs trilinear coupling is extremely challenging at the LHC, given the small rate and interference with other Standard Model contributions in the same channels, and the anticipated measurement is of order $100\%$, which loosely corresponds to a bound $c_6 \frac{v^2}{\Lambda^2} \lesssim 0.065$. 

In contrast, linear colliders are extraordinary environments in which to measure Higgs self-interactions. At the ILC, the latest projections for direct measurement of the Higgs self-coupling achieve 26.6\% precision assuming a Standard Model-like value \cite{duerig}. This corresponds to a bound on corrections to the Standard Model prediction of
 \begin{equation}
 c_6 \frac{v^2}{\Lambda^2} < 0.017
 \end{equation}
 which for $c_6 = 1$ corresponds to a scale $\Lambda > 1.3$ TeV. 
 
The argument for circular colliders is a bit different, assuming they are operating below threshold for directly production of Higgs pairs. At circular lepton colliders such CEPC with a center-of-mass energy in the vicinity of 250 GeV, a direct measurement of the self-coupling is not possible. However, corrections to the Higgs self-coupling coming from $\mathcal{O}_6$ will appear in radiative corrections to other observables measured with high precision, particularly the $Zh$ cross section \cite{McCullough:2013rea}. Assuming no other deviations from the Standard Model, a precision measurement of the $Zh$ cross section would yield a bound of
  \begin{equation}
 c_6 \frac{v^2}{\Lambda^2} < 0.023
 \end{equation}
 corresponding to a scale $\Lambda > 1.1$ TeV (for $c_6 = 1$).
 
 Of course, we do not generally expect to see departures from the Standard Model in the form of individual irrelevant operators appearing in isolation. If generic departures from the Standard Model involve a combination of irrelevant operators with various Wilson coefficients, the interpretation of these precision measurements is altered. This is often a criticism leveled at the measurement of the Higgs self-coupling via corrections to the $Zh$ cross section at circular lepton colliders, but in reality it applies to {\it any} measurement of the Higgs self-coupling, whether or not it involves pairs of Higgs bosons in the final state. All such measurements accumulate corrections from an array of operators, and any bound on generic new physics requires assembling a variety of measurements to constrain the space of possible deviations. 
 
 While this is a complicated exercise in general, it is useful to focus on particularly motivated classes of new physics. These scenarios single out subspaces in the space of possible corrections, and allow for a relatively straightforward assessment of the precision achievable at lepton colliders. A particularly useful way to define motivated classes of new physics is by their Standard Model quantum numbers. Such new physics can be charged under any combination of the $SU(3)_c \times SU(2)_L \times U(1)_Y$ forces of the Standard Model, or may be entirely neutral under the Standard Model. This latter possibility is particularly compelling, as new physics entirely neutral under the Standard Model can only couple to us through one of two possible relevant or marginal interactions: kinetic mixing with hypercharge of the form $F_{\mu \nu} \mathcal{O}^{\mu \nu}$, and the Higgs portal coupling $\mathcal{O}|H|^2$. In the latter case, we can think of these SM-neutral states as being charged under the ``Higgs force'', as the Higgs mediates a Yukawa force between such particles. While the strength of the Higgs force is infinitesimally small compared to the other Standard Model forces, for particles neutral under the Standard Model it may be the leading interaction. 
 
In general, heavy states interacting with the Standard Model exclusively through the Higgs force could not be meaningfully probed prior to the discovery of the Higgs, and in this respect populate an entirely-unknown frontier. With the discovery of the Higgs, however, we can explore the Higgs frontier for the first time, and the precision of lepton colliders is an exceptional tool.
   
 \subsection{Probing the Higgs Force}
 
Measuring the ``size'' of the Higgs and the strength of its self-interactions turns out to be deeply connected to probing new physics that is entirely neutral under the Standard Model, a task for which future lepton colliders are ideally suited. Integrating out heavy states that are singlets under the Standard Model gauge interactions and couple strictly through the Higgs portal $|H|^2$ generates only two operators, namely the operators $\mathcal{O}_H$ and $\mathcal{O}_6$ discussed above. Thus SM-neutral particles coupling through the Higgs force define a particularly motivated subspace in the space of possible irrelevant operators correcting the Standard Model. Amusingly, these operators are precisely the ones we have associated with a finite ``size'' of the Higgs and deviations in the Higgs self-coupling.

\begin{figure}[t]    \centering
   \includegraphics[width=4in]{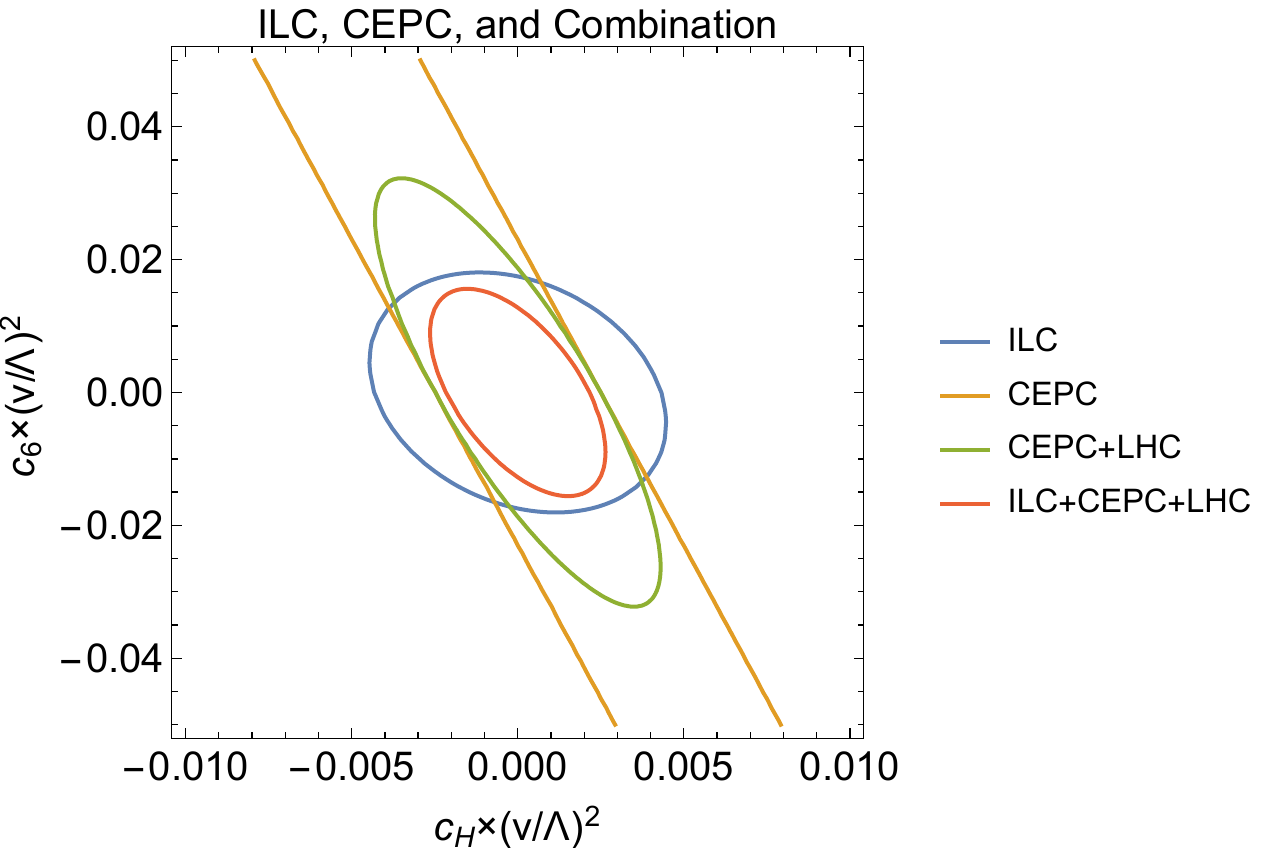} 
   \caption{Bounds on corrections to the Standard Model from SM-singlet new physics, parameterized by the irrelevant operators $\mathcal{O}_H$ and $\mathcal{O}_6$. Here the limits are $1\sigma$ bounds coming from the ILC measurement of $Zh$ and $Zhh$ cross sections (blue), CEPC measurement of the $Zh$ cross section (orange), the CEPC measurement in conjunction with a 50\% measurement of the Higgs self-coupling at the LHC (green), and the combination of ILC, CEPC, and LHC measurements (red). This makes particularly clear the sense in which measurements below the $Zhh$ threshold can be used to constrain deviations in the Higgs self-coupling in the presence of new physics contributing to multiple effective operators.}
   \label{fig:combofit}
\end{figure}

We can now meaningfully ask how the two-dimensional space of $c_H$ and $c_6$ can be constrained by proposed linear and circular colliders. In the case of circular colliders, the dominant precision comes from the measurement of the $Zh$ cross section, which can be used to constrain a linear combination of $c_H$ and $c_6$.\footnote{Additional precision is, of course, achievable using a combination of all measurements, but given that the precision at circular colliders is driven by the precision in $Zh$, these corrections will be relatively weak.} Linear colliders operating above the $Zhh$ threshold contribute measurements of both $Zh$ and $Zhh$ rates, each of which accumulate contributions proportional to $c_H$ and $c_6$.\footnote{Here I am using the results from \cite{McCullough:2013rea, Tanabe} to obtain the covariance matrix.} Neglecting correlated errors, we can place corresponding constraints in the plane of $c_H$ and $c_6$, as shown in Figure \ref{fig:combofit}. 

The achievable precision is evidently considerable, but the physics value of these limits is more transparent when phrased in terms of the masses and couplings of new particles. A particularly useful example for translating this precision into bounds on the masses and couplings of SM-neutral particles is the case of a real singlet scalar field $\Phi$ coupling to the Standard Model via Higgs portal interactions. For completeness, allowing for self-interactions of this new singlet scalar, the most general renormalizable Lagrangian is
\begin{equation}
\mathcal{L} = \frac{1}{2} \left(\partial_\mu \Phi \right)^2 - \frac{1}{2} m^2 \Phi^2 - A |H|^2 \Phi - \frac{1}{2} k |H|^2 \Phi^2 - \frac{1}{6} \mu \Phi^3 - \frac{1}{24} \lambda \Phi^4
\end{equation}
The Wilson coefficients for $\mathcal{O}_H$ and $\mathcal{O}_6$ generated by integrating out a heavy $\Phi$ arise at both tree-level and one loop \cite{Henning:2014wua},
\begin{eqnarray}
\Delta \mathcal{L}_{\rm tree} &=& \frac{A^2}{m^4} \mathcal{O}_H + \left( - \frac{k A^2}{2m^4} + \frac{1}{6} \frac{\mu A^3}{m^6} \right) \mathcal{O}_6 \\
\Delta \mathcal{L}_{\rm loop} &=& \frac{1}{16 \pi^2} \frac{1}{m^2} \left( \frac{k^2}{12} \mathcal{O}_H - \frac{k^3}{12} \mathcal{O}_6 \right)
\end{eqnarray}
For nonzero $A$, there are generally quite a few ways to probe $\Phi$; electroweak symmetry breaking leads to tree-level mixing between $\Phi$ and $H$, leading to large corrections to Higgs couplings and allowing direct searches for resonant production of the mostly-$\Phi$ mass eigenstate. However, for $A= 0$ (consistent with, e.g., a $\mathbb{Z}_2$ symmetry under which $\Phi \rightarrow - \Phi$) the tree-level corrections vanish, there is no resonant production mode for $\Phi$, and only subtle loop-level corrections remain. 

\begin{figure}[t]    \centering
   \includegraphics[width=4in]{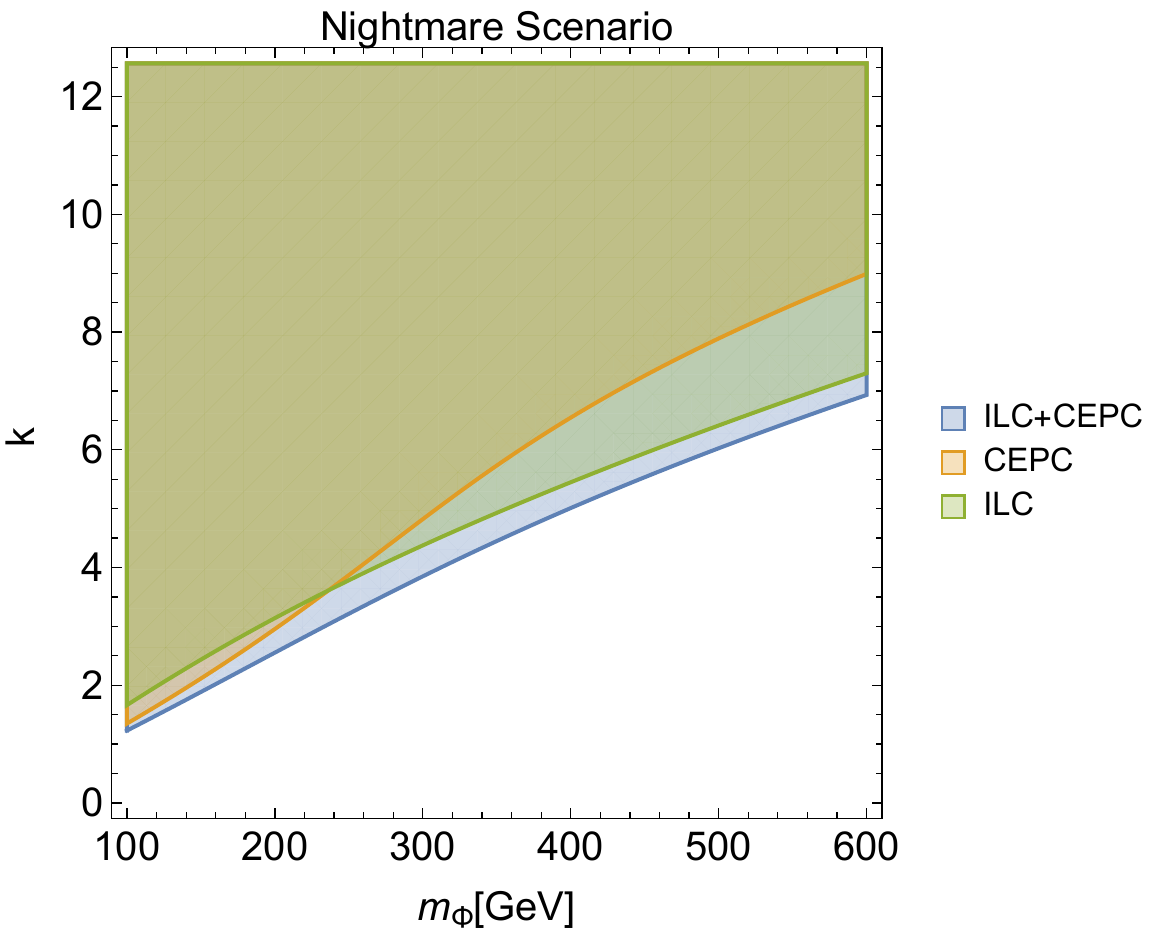} 
   \caption{$1 \sigma$ constraints on the mass of a real singlet $m_\Phi$ and the strength of the Higgs portal coupling $k$ from measurements of the $Zh$ cross section at the CEPC, $Zh$ and $ZZh$ cross sections at the ILC, and the combination thereof. The crossover in sensitivity between linear and circular colliders is due to the relative powers of $k$ appearing in corrections to $Zh$ and $ZZh$ rates.  These bounds are computed in the EFT limit, which breaks down at low $m_\Phi$, but agreement with the full result \cite{Englert:2013tya, Craig:2013xia,Curtin:2014jma} is quantitatively good for $m_\Phi \gtrsim 200$ GeV and qualitatively good throughout the entire mass range.}
   \label{fig:nightmare}
\end{figure}

Remarkably, as shown in Figure \ref{fig:nightmare}, this ``nightmare scenario'' can still be powerfully constrained by lepton colliders. In particular, the ability of lepton colliders to constrain the nightmare scenario arises from the extraordinary precision of the $Zh$ coupling measurement, such that it provides sensitivity to loop-level corrections from SM-neutral new physics to tree-level observables in the Standard Model \cite{Englert:2013tya}. Although the sensitivity is primarily restricted to relatively large Higgs portal couplings, such couplings are precisely those motivated by considerations of electroweak naturalness \cite{Craig:2013xia} and electroweak baryogenesis \cite{Curtin:2014jma}.

\section{Lepton Colliders as Machines for Exploration}

Lepton colliders are typically thought of as machines for precision measurement rather than direct discovery, but the clean environment relative to hadron colliders provides several particularly sharp opportunities for exploring the unknown through the production of new particles. For the purposes of this section I'll focus on the discovery potential of the ILC, since the modest increase in center-of-mass energy of CEPC relative to LEP makes for a comparatively narrower window for discovery. The arguments here will also be somewhat more qualitative in nature, though excellent quantitative arguments have been made in e.g. \cite{Fujii:2015jha, Fujii:2017ekh}.

The essential point is that, while the LHC is a fantastic machine for exploration, its strengths are greatest for exploring new physics charged under the strong interactions, for which abundant production rates compensate for the messiness of hadronic processes. By the end of the LHC, we will have rather comprehensively probed the existence of colored particles to the TeV scale and somewhat beyond.  But this leaves considerable room to search for particles charged under more subtle forces, particularly the weak force or the Higgs force discussed earlier. Not coincidentally, these forces are the most relevant to dark matter and other motivated candidates for kinematically-accessible physics beyond the Standard Model. In any event, the discovery potential of lepton colliders is not restricted to dark matter alone; very generally they are ideal tools for exploring any new physics coupling through the weak or Higgs forces.

\subsection{Exploring the Weak Frontier}

\begin{figure}[t]    \centering
   \includegraphics[width=3in]{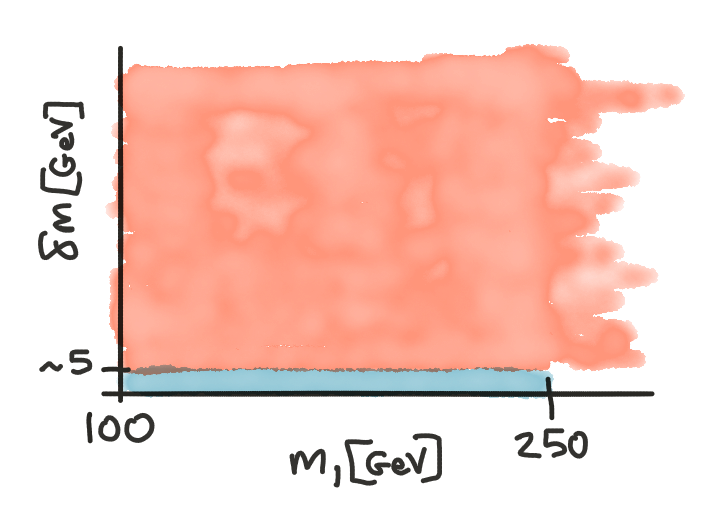} 
   \caption{A sketch of LHC and ILC reach for fermions carrying electroweak quantum numbers in terms of the mass $m_1$ of the lightest mass eigenstate (assumed to be charge neutral) and the mass splitting $\delta m$ to the next-to-lightest mass eigenstate. Here the LHC coverage is in red, and the unique ILC coverage (i.e., regions in which the ILC could set new limits) is in blue. }
   \label{fig:ewkold}
\end{figure}

It is tempting to argue that the LHC will comprehensively cover kinematically accessible new physics carrying electroweak quantum numbers. After all, the LHC places visually impressive limits on electroweakinos. However, this impressive reach is a bit misleading, in that it relies on large splittings between states in the same electroweak multiplet, at least in the event that the lightest state is electromagnetically neutral.\footnote{New electroweak multiplets whose lightest component is charged and stable on even modestly macroscopic length scales can, of course, be comprehensively excluded by the LHC to the TeV scale.}  Such splittings lead to large amounts of missing energy, compensating for relatively small electroweak cross-sections. This is a sensible expectation given the historical way of thinking about weak-scale supersymmetry, in which we expected an abundance of new states around the same scale with sizable mixings. Then it makes sense to plot limits in terms of, say, the mass of the lightest electroweakino and the mass splitting to the next state. In this plane, the LHC covers a broad range of masses and mixings, and only loses sensitivity when splittings grow small and missing energy is diminished. Plotted this way, the LHC reach (and the corresponding potential for the ILC to improve limits relative to the LHC) for electroweakinos looks something like the sketch Figure \ref{fig:ewkold}. This corresponds to the inability of the LHC to probe e.g. pure Higgsinos with small mass splittings in the Higgsino multiplet, which are a key target for the ILC  \cite{Fujii:2017ekh}.

But thus far we have found little evidence for an abundance of new states near the weak scale. In that case, the sensible way to parameterize the sensitivity of colliders to new states charged under the electroweak interactions is to consider various possible electroweak representations one at a time, and investigate the natural parameter space dictated by symmetries. If we just imagine adding one new set of electroweak fermions to the Standard Model, at tree level the fermions will be degenerate because there is a symmetry -- electroweak symmetry! -- that guarantees their degeneracy. Any splittings will then come from electroweak symmetry breaking. Standard Model interactions alone generate these splittings at loop level, while if there is mixing with a heavier state, the splittings will parametrically be of the form $v^2/\Lambda$, where $\Lambda$ is the scale of the heavier state (typically some additional electroweak representations mixing with the light fermions). Even the radiative Standard Model splittings induced by electroweak symmetry breaking in a given multiplet can be expressed in this language, and are typically of order $\alpha m_Z$ -- so in the absence of further additional fermions, the splittings induced via electroweak symmetry breaking correspond to $\Lambda \sim 4 \pi v / (e^2 g_z^2)$.

\begin{figure}[t]    \centering
   \includegraphics[width=3in]{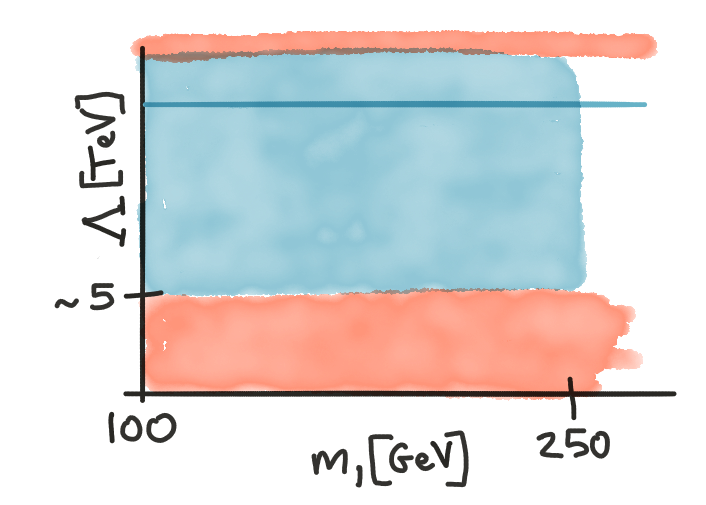} 
   \caption{A sketch of LHC and ILC reach for fermions carrying electroweak quantum numbers in terms of the mass $m_1$ of the lightest mass eigenstate (assumed to be charge neutral) and the mass scale $\Lambda$ of additional physics leading to splitting in the specified electroweak multiplet. Again the LHC coverage is in red, and the unique ILC coverage is in blue. The solid blue line denotes the effective scale $\Lambda$ corresponding to irreducible splittings induced at one loop from electroweak symmetry breaking. The location of this line depends on the particular electroweak quantum numbers of the fermion multiplet. LHC limits reappear as $\Lambda \rightarrow \infty$ due to searches for heavy stable charged particles and disappearing charged tracks.}
   \label{fig:ewknew}
\end{figure}

In this parameterization of new electroweak multiplets beyond the Standard Model without large mixing, the more natural parameter space in which to plot limits is in terms of the mass of the new multiplet versus the scale of induced splittings $\Lambda$. This clearly parameterizes the higher-dimensional operators that induce splitting in the multiplet after EWSB, and it correctly organizes the problem around the symmetries -- namely electroweak symmetry. When the parameter space is plotted in this way, the LHC and ILC reach for new electroweak fermions takes the schematic form of Figure \ref{fig:ewknew}. In terms of this parameterization, it's clear that a lepton collider with high center-of-mass energy is covering a significant region of the physically sensible parameter space, while the LHC covers the narrower cases where there is significant mixing or the splitting is tuned to be unnaturally small (i.e. below the irreducible Standard Model splitting).

Given that there is not an apparent abundance of new physics at the weak scale, the appropriate way to parameterize searches for new physics charged under the weak interactions is to consider adding one state at a time and parameterizing the space of masses and splittings accordingly. Phrased in these terms, lepton colliders with high center-of-mass energies such as the ILC are outstanding machines for probing new physics charged under the weak interactions.

\subsection{Exploring the Higgs Frontier}

\begin{figure}[t]    \centering
   \includegraphics[width=5in]{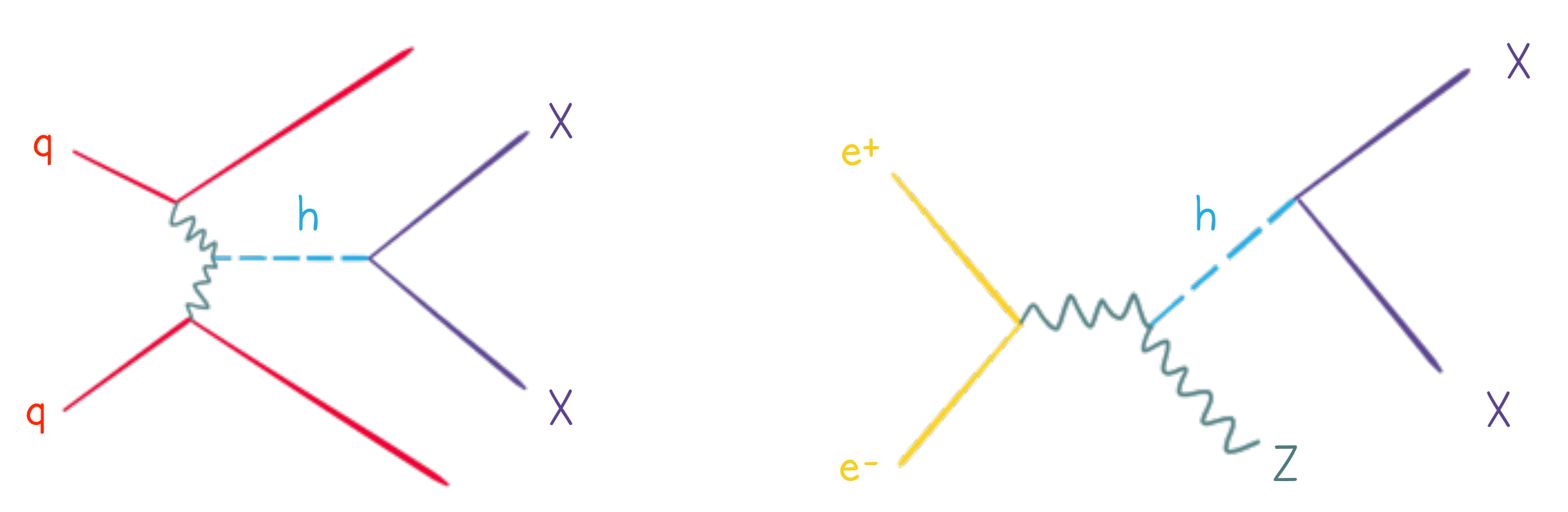} 
   \caption{Production modes at the LHC (left) and lepton colliders (right) for new particles coupling to the Standard Model purely through the Higgs portal.}
   \label{fig:offshell}
\end{figure}

Unsurprisingly, lepton colliders are also excellent machines for exploring new physics that is entirely neutral under the Standard Model and couples only via the Higgs force. We have already seen the sense in which precision Higgs coupling measurements can be used to constrain heavy states whose effects can be parameterized in terms of effective operators involving Standard Model fields. But lepton colliders are also powerful machines for directly discovering such states if they are light enough to be produced on shell. A canonical example is that of scalars or fermions that are neutral under the Standard Model and couple only through the Higgs portal. When these particles are lighter than half of the Higgs mass, they may be produced in decays of the Higgs boson and constrained by direct measurements of Higgs invisible decays. When they are heavier than half of the Higgs mass, they may nonetheless be produced via off-shell Higgses at the LHC and ILC in the processes illustrated in Figure \ref{fig:offshell}. The rate for this off-shell process is too small to be effectively constrained by the LHC, but may be meaningfully constrained by direct searches at the ILC and other lepton colliders with a high center-of-mass energy. 

\begin{figure}[t]    \centering
   \includegraphics[width=4in]{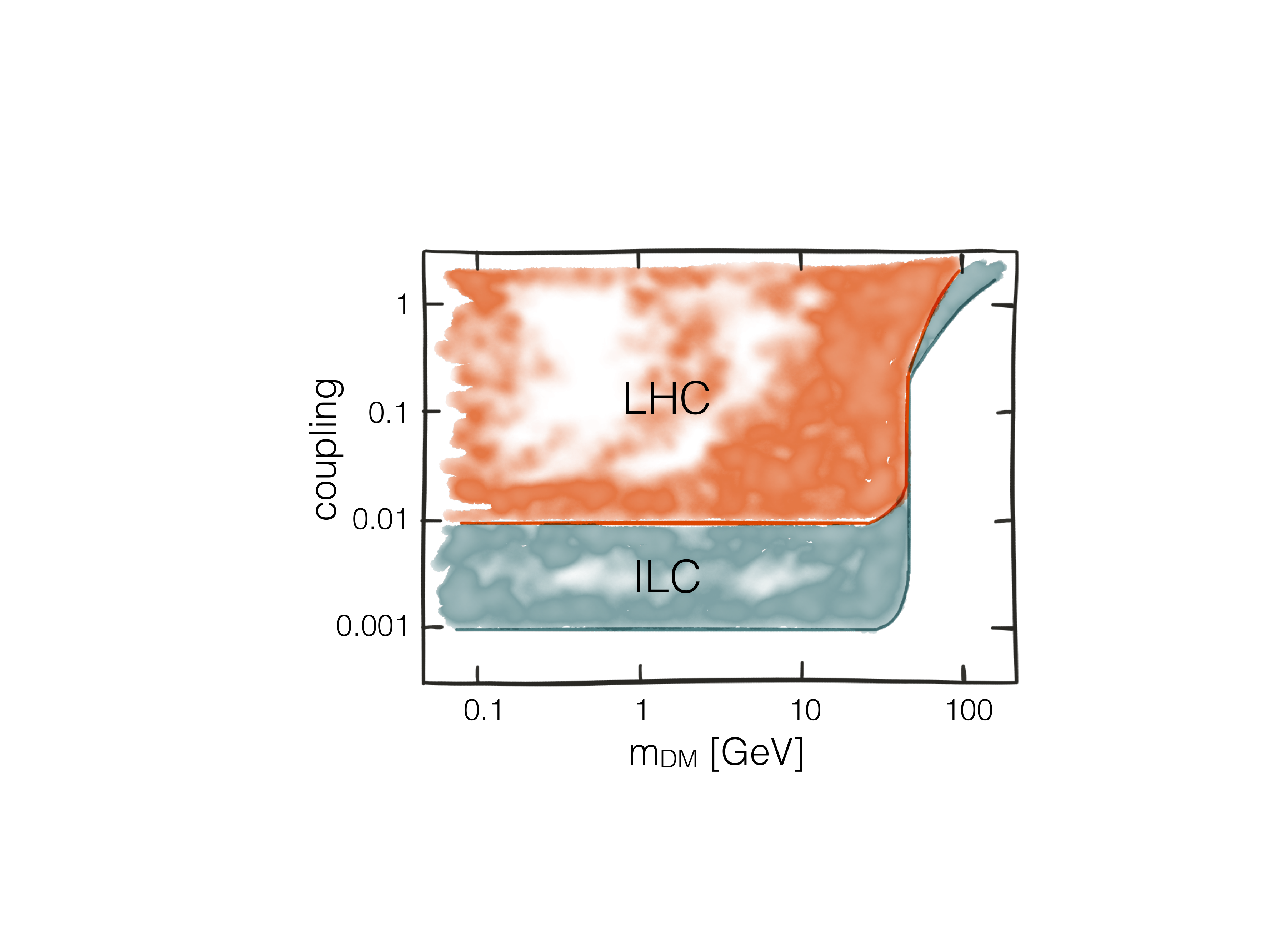} 
   \caption{Cartoon of the LHC and ILC reach for direct pair production of new SM-neutral states produced via the Higgs portal in terms of the dimensionless Higgs portal coupling and mass of the new particle. See e.g. \cite{Chacko:2013lna, Craig:2014lda}.}
   \label{fig:direct}
\end{figure}

The relative reach of LHC and ILC  searches for the on-shell and off-shell Higgs-mediated pair production processes is sketched in Figure \ref{fig:direct} using the expected limits on Higgs invisible decays in conjunction with results in e.g. \cite{Chacko:2013lna, Craig:2014lda}. Apart from considerably extending the reach of searches for states coupling via the Higgs portal, the projected ILC reach makes significant inroads into regions relevant for Higgs portal dark matter. This makes clear the sense in which lepton colliders are ideally suited for probing new physics charged solely under the ``Higgs force''.

\section{Conclusions}

A wide range of arguments may be made in favor of future particle accelerators. Here I have attempted to convey the sense in which lepton colliders are ideal machines for future understanding and exploration. As machines for understanding, they provide unprecedented tools for studying the Higgs and observing phenomena never before seen in nature. As machines for exploration, they are ideally suited to discovering new particles interacting too weakly to be seen otherwise. These arguments are largely independent of the outcome of the LHC physics program, and provide compelling rationale for future linear and circular lepton colliders.

\section*{Acknowledgements} I am grateful to the participants and organizers of the International Workshop on Future Linear Colliders (LCWS2016) in Morioka, Japan for the opportunity to articulate these ideas. I owe considerable thanks to Nima Arkani-Hamed, Tim Barklow, Marco Farina, Jiayin Gu, Zhen Liu, Matthew McCullough, Maxim Perelstein, Michael Peskin, Kechen Wang, and Liantao Wang for illuminating conversations about the physics potential of future lepton colliders. I am especially grateful to Nima Arkani-Hamed for the encouragement to think big. This work is supported in part by the US Department of Energy under the grant DE-SC0014129.


\bibliography{leptonrefs}
\bibliographystyle{utphys}

\end{document}